\documentclass{article}

\usepackage{microtype}
\usepackage{graphicx}
\usepackage{subcaption}
\usepackage{booktabs}
\usepackage{enumitem}
\usepackage{hyperref}

\usepackage[preprint]{icml2026}

\usepackage{amsmath}
\usepackage{amssymb}
\usepackage{mathtools}
\usepackage{amsthm}

\usepackage[capitalize,noabbrev]{cleveref}

\definecolor{rulegray}{gray}{0.72}
\newcommand{\nvidialetterhead}{%
  \noindent\includegraphics[height=0.5in]{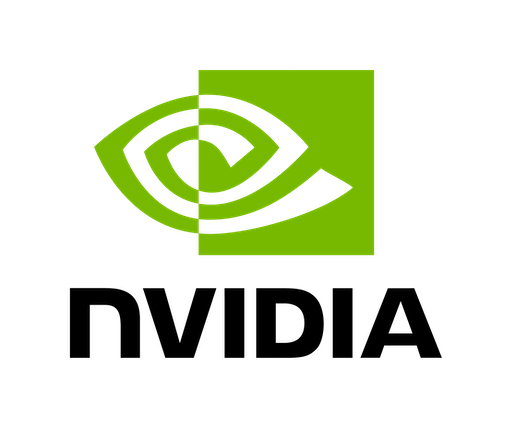}\par
  \vskip 3pt
  \noindent{\color{rulegray}\rule{\textwidth}{0.7pt}}%
  \vskip 6pt
}

\theoremstyle{plain}

\theoremstyle{definition}

\icmltitlerunning{AEVAL: Deterministic Testing for Agentic Skill Workflows}

\makeatletter
\gdef\icmlcorrespondingauthor@text{Tian Zheng \textless{}tiazheng@nvidia.com\textgreater{}, Tejas Singh Anand \textless{}tejassingha@nvidia.com\textgreater{}, Yuet Ying Christina Wang \textless{}christwang@nvidia.com\textgreater{}}
\renewcommand{\Notice@String}{}
\makeatother

\begin{document}

\twocolumn[
  \nvidialetterhead

  \icmltitle{AEVAL: From Anecdotal to Deterministic Testing \\
    for Agentic Skill Workflows}

  \icmlsetsymbol{intern}{*}

  \begin{icmlauthorlist}
    \icmlauthor{Tejas Singh Anand}{intern,nvidia,uw}
    \icmlauthor{Yuet Ying Christina Wang}{nvidia}
    \icmlauthor{Wanting Jiang}{nvidia}
    \icmlauthor{Steve Masson}{nvidia}
    \icmlauthor{Tian Zheng}{nvidia}
    \icmlauthor{Bingjie Zhou}{nvidia}
  \end{icmlauthorlist}

  \icmlaffiliation{nvidia}{NVIDIA}
  \icmlaffiliation{uw}{University of Waterloo}

  \icmlkeywords{Agentic AI, Evaluation, Uncertainty, Self-Correction Bias, Conformal-style Guarantees, LLM Agents}

  \vskip 0.3in
]

\printAffiliationsAndNotice{\textsuperscript{*}Tejas Singh Anand contributed to this work during an internship at NVIDIA.}

\begin{abstract}
Modern agentic systems increasingly rely on \emph{skills}: installable packages of natural language and code that teach an LLM agent how to perform a domain task. As skill repositories grow, developers need automated, trustworthy quality signals on every change to a skill, yet most evaluation today is \emph{anecdotal}: a developer asks an agent to ``try the skill,'' watches a demo turn, and forms a subjective impression of whether it works. This produces neither reproducibility across runs nor comparability across versions, and it scales poorly to multi-skill marketplaces where a single regression can silently break dozens of downstream workflows. We present \textbf{AEVAL} (\emph{Agentic Evaluation}), a CI-integrated framework that replaces this anecdotal practice with a deterministic, reproducible test pipeline for agentic skills. The framework treats every skill change as a triggered test event, runs each skill against a developer-declared evaluation contract (\texttt{eval.config}) inside an automated executor, and emits a structured, evidence-grounded quality signal that downstream CI can route on. A key technical ingredient is a structural separation between the executor and the grader, which prevents a subtle but pervasive failure mode of agentic evaluation: an agent that silently self-corrects during execution and then grades its own patched outputs as passing. Our contributions are: (i) a deterministic, change-triggered evaluation protocol for skills with per-skill evaluation contracts and per-run artifact schemas; (ii) a formalization of \emph{self-correction bias} as a distinct failure mode of naive agentic evaluators; (iii) a structural executor/grader separation with a first-attempt grading rule and explicit self-correction tracking; and (iv) a tiered, grounded-evidence suggestion scheme (LV1 causal fixes, LV2 quality improvements) posted as inline merge-request comments. We validate the framework against a runtime-agnostic interface tested with multiple popular agent SDKs. Across a set of real skills in a production agentic stack, the protocol converts spurious 100\% pass rates into reproducible first-attempt FAIL signals with an auditable record of every fix the executor applied, enabling reliable routing and stopping decisions in downstream agentic workflows.
\end{abstract}

\section{Introduction}
\label{sec:intro}

Agentic coding systems increasingly externalize domain capability through \emph{skills}: small, versioned packages containing a natural-language specification (typically a \texttt{SKILL.md}), supporting scripts, and configuration~\cite{anthropic2024skills}. An agent runtime discovers a skill from its library and follows its instructions to perform a task. In a production stack, a skill failure is effectively a product failure: a regression in a single skill file silently breaks every downstream workflow that depends on it.

Despite this, evaluation of skills today is overwhelmingly \emph{anecdotal}. The dominant practice is for a developer to open an interactive session, paste a representative prompt, observe a demo run, and form a subjective impression of whether the skill ``still works.'' The signal is informal, non-reproducible, and incomparable across versions: two consecutive runs may exercise different code paths, produce different outputs, and elicit different developer judgments. As a skill marketplace grows to dozens or hundreds of artifacts, this practice does not scale, and silent regressions accumulate between releases. We argue that what is needed is the same shift that software engineering underwent decades ago: from anecdotal manual testing to a deterministic, change-triggered test pipeline whose outputs are reproducible artifacts rather than human impressions.

This paper introduces \textbf{AEVAL}, a framework that performs that shift for agentic skills. AEVAL treats every change to a skill as a CI-triggered test event. The skill ships with an explicit \emph{evaluation contract} (\texttt{eval.config}) declaring its test prompt, expected outcome, and required credentials; the framework installs the modified skill into a clean session, executes it against the contract via an automated agent runtime, and emits a structured, machine-readable quality signal (per-assertion pass/fail with cited evidence, a first-attempt pass rate, an analysis pass over benchmark statistics, and tiered fix suggestions). The signal is reproducible across runs, comparable across MRs, and consumable by downstream CI without human interpretation. In practice this turns the \texttt{eval.config} into a \emph{persistent, declarative test harness}: the developer describes the test workflow once at skill-creation time, and every subsequent push automatically replays the full install--execute--grade--suggest-fixes loop with no human in the loop. There is no need to reopen an interactive coding agent, re-load project context, or re-supply credentials per run; the harness is a checked-in artifact that any push or manual CI trigger replays end-to-end. When grading fails, AEVAL also acts as the developer's automated reviewer: it posts inline fix-suggestion commits on exactly the source lines responsible for the failure, ready for the original author to apply with a single click and re-trigger the pipeline.

Building a deterministic pipeline on top of a stochastic agent introduces a specific reliability problem that any naive instantiation will inherit, and that we address as a core technical contribution. When the same agent both executes and grades the skill, the signal is systematically optimistic: LLM agents are \emph{trained} to recover from errors~\cite{saunders2022self,bai2022constitutional} and will silently retry, amend configuration, or patch the skill in place until something succeeds, then truthfully report the state of the world they produced. The resulting evaluation measures agent debugging ability rather than skill quality, a pattern closely related to observed self-preference effects in LLM-as-judge settings~\cite{panickssery2024llm,zheng2023judging}. We refer to this failure mode as \emph{self-correction bias}, formalize it in \cref{sec:bias}, and prevent it through a structural separation between the executor and the grader (\cref{sec:separation}). Without this separation, the deterministic pipeline returns reproducible \emph{noise}; with it, it returns a reliable signal that downstream uncertainty quantification methods~\cite{kadavath2022know,lin2022teaching,angelopoulos2023ppi} can consume.

\paragraph{Contributions.} We make four contributions.
\begin{enumerate}[leftmargin=*,itemsep=1pt,topsep=2pt]
    \item \textbf{Deterministic skill testing} (\cref{sec:method,sec:system}): a change-triggered evaluation protocol for agentic skills with per-skill evaluation contracts, per-run artifact schemas, and an agent-runtime-agnostic executor interface, replacing anecdotal manual evaluation with a reproducible CI pipeline.
    \item \textbf{Self-correction bias} (\cref{sec:bias}): we formalize a distinct failure mode of agentic evaluators and show it arises structurally rather than from a specific model choice.
    \item \textbf{Structural separation and first-attempt grading} (\cref{sec:separation}): a protocol that constrains the grader to a restricted information set (outputs plus execution transcript) so it cannot participate in corrective action. Combined with a \emph{first-attempt grading rule} and explicit self-correction tracking, this yields a quality signal grounded in the skill as shipped rather than the skill as patched.
    \item \textbf{Grounded, tiered suggestions delivered as one-click MR commits} (\cref{sec:suggestions}): every suggested fix must cite a specific failed assertion or tracked self-correction entry, with LV1 (causally implicated in failure) and LV2 (improvements flagged by the grader) tiers preventing speculative ``nice to have'' churn. Each suggestion is posted as an inline GitLab MR comment with an ``Apply suggestion'' diff, so when AEVAL detects a defect it also produces an empirically validated patch the original developer can apply in a single click, closing the test-fail-fix loop within the same merge request and turning the framework into an automated reviewer for the skill author.
\end{enumerate}

\section{Background and Related Work}
\label{sec:related}

\paragraph{Agent evaluation frameworks.} Existing platforms sit at one of three layers. \emph{Prompt/response} platforms (LangSmith~\cite{langsmith2025}, Braintrust~\cite{braintrust2025}, Promptfoo~\cite{promptfoo2025}) score model outputs against assertions; they do not install or execute deployable skill artifacts. \emph{Capability benchmarks} (OpenAI Evals~\cite{openai2024evals}, AgentBench~\cite{liu2024agentbench}, SWE-bench~\cite{jimenez2023swebench}) evaluate a \emph{model's} general problem-solving on a fixed dataset; they cannot be pointed at a developer's modified skill in an MR. \emph{Sandboxed agent evaluation} (Harbor Framework~\cite{harbor2025}) runs agents against predefined tasks in containers; it tests the agent, not the skill artifact, and does not spawn a structurally independent evaluator.

\paragraph{LLM-as-judge and self-evaluation.} LLM-as-judge protocols~\cite{zheng2023judging} and self-evaluation pipelines~\cite{saunders2022self,ren2023selfeval} are known to exhibit self-preference~\cite{panickssery2024llm} and calibration issues~\cite{kadavath2022know,lin2022teaching}. These works study judgment \emph{of model outputs}. Our setting is different: we study judgment of a \emph{skill artifact's behavior} under an executor that actively modifies the world during evaluation.

\paragraph{Uncertainty quantification for agents.} Distribution-free methods such as conformal prediction~\cite{vovk2005algorithmic,angelopoulos2023gentle} and prediction-powered inference~\cite{angelopoulos2023ppi} provide rigorous coverage under exchangeability. Our protocol is complementary: we do not compete with conformal guarantees but rather prepare a \emph{reliable ground-truth signal} that such methods could consume. Without self-correction-bias-free evaluation, downstream conformal or sequential procedures are fed contaminated labels and their nominal guarantees collapse.

\section{The Self-Correction Bias Problem}
\label{sec:bias}

Let $s$ be a skill, $x$ an eval prompt from a developer-supplied \texttt{eval.config}, and $A$ an autonomous agent. A naive evaluator runs $A$ on $(s, x)$, receives a terminal state $y$, and emits $\hat{q}(s) \in \{0,1\}$ indicating whether $s$ worked. We observe that $A$'s execution trajectory
\begin{equation*}
  \tau = (a_1, o_1, a_2, o_2, \dots, a_T, o_T)
\end{equation*}
is typically not a read-only probe of $s$. Because $A$ has tools (edit, bash, write), actions $a_t$ may modify the skill itself or adjacent configuration. Let $\Delta(\tau)$ denote the set of skill-file edits executed during $\tau$. Define
\begin{align*}
  q^\star(s) &= \mathbb{I}[s \text{ works first-attempt, unmodified}] \\
  \hat{q}_{\text{naive}}(s) &= \mathbb{I}[y \text{ satisfies } x]
\end{align*}

The failure mode is that $\hat{q}_{\text{naive}}(s) = 1$ whenever $A$ can recover via any $\Delta(\tau) \neq \emptyset$, even when $q^\star(s) = 0$. This is structural: an agent that is competent at debugging will hide defects in any skill $s$ that is merely \emph{fixable} rather than \emph{broken}. We observed this empirically in our own deployment, an agent-first product in which skills are authored by developers and reach end users without an intermediate packaging step. In this setting every developer push to a skill is itself a production event, so a regression masked by a self-correcting evaluator ships directly. On one such push, a first-run execution failed with a missing configuration entry, the agent silently patched the skill in place, re-ran, and produced a 100\% pass grade against assertions it had itself authored after observing the corrected output. This is exactly the bias the formal model above predicts.

Two design forces cause this. First, \emph{shared identity}: when the same agent runs and grades, it has no incentive to flag its own patches. Second, \emph{post-hoc assertions}: assertions written after seeing outputs implicitly fit the outputs the agent produced. Eliminating the bias requires addressing both.

\section{Method}
\label{sec:method}

\begin{figure}[t]
\centering
\includegraphics[width=0.95\columnwidth]{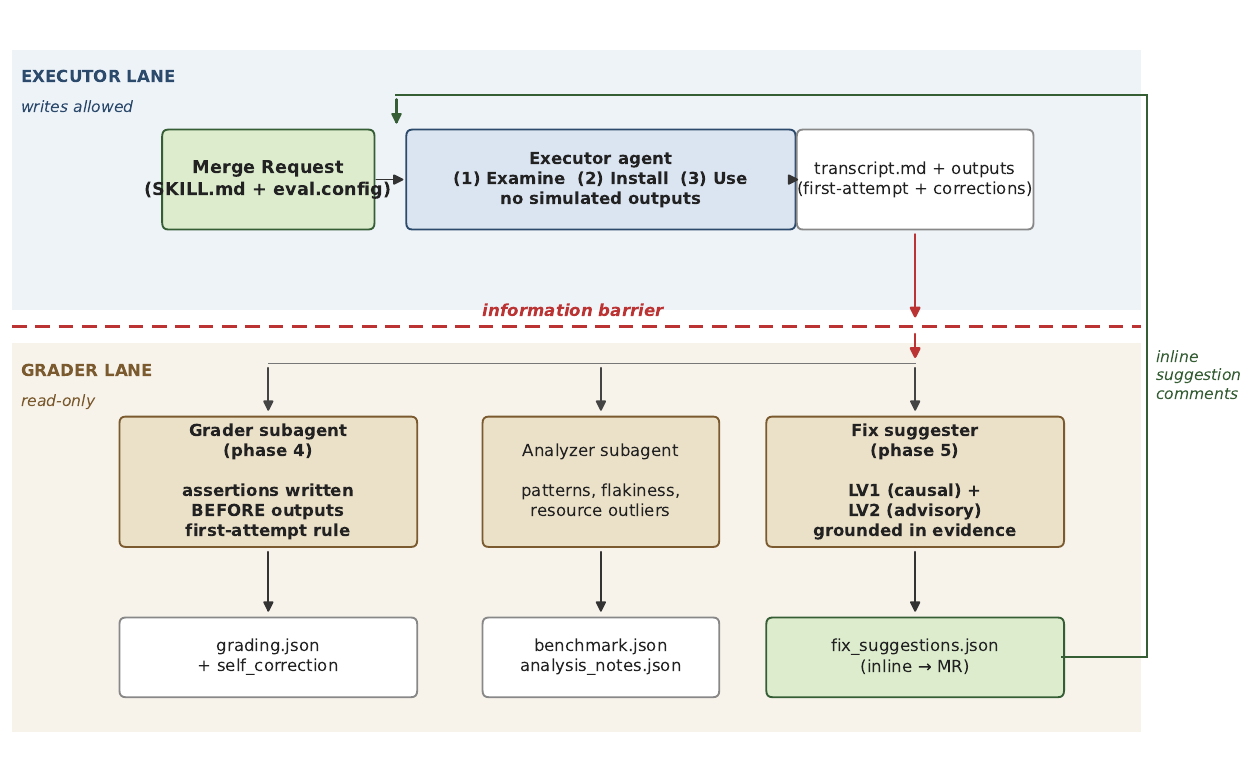}
\caption{AEVAL pipeline. An MR triggers CI, which detects changed skills and invokes the evaluator. The executor runs the four-phase workflow; the grader is a separate subagent with access only to outputs and transcript; suggestions are posted to the MR.}
\label{fig:pipeline}
\end{figure}

AEVAL is a five-phase pipeline triggered by a merge request on a skill directory (\cref{fig:pipeline}). The core pipeline runs: \textbf{(1) Examine}, \textbf{(2) Install}, \textbf{(3) Use}, \textbf{(4) Evaluate}, and \textbf{(5) Suggest}. The distinctive design choices sit in phases 3--5.

\subsection{Skill-Defined Evaluation Contract}
\label{sec:evalconfig}

Each skill optionally ships an \texttt{eval.config} file co-located with its \texttt{SKILL.md}. The contract specifies, per test case, a natural-language \emph{prompt}, an \emph{expected outcome}, and a list of \emph{required credentials}. Multiple independent cases are supported via an \texttt{evals} array, each executed in an isolated SDK session. A three-tier fallback (per-skill, per-name, generic) ensures every skill is testable even without explicit configuration. Locating the test specification with the skill ensures each artifact declares its own quality criteria, moving the eval contract into the MR's diff and out of a centralized benchmark registry.

\subsection{Banning Simulated Outputs}

Phase 3 of the executor prompt imposes a hard rule: \emph{no simulated outputs}. If a command fails, a service is unreachable, or a workflow errors, that failure is recorded in a per-case transcript $\tau$ as a real test result. The agent may self-correct, but it cannot fabricate logs, synthesize results, or place-hold missing outputs. This is enforced by instruction rather than sandbox, but we found it essential: without it, agents routinely produce synthetic training logs when a GPU cluster was unreachable and grade those as passing.

\subsection{Separating Execution from Grading}
\label{sec:separation}

Phase 4 grades the skill. The core constraint is that the \emph{grader is a separate subagent}, spawned from a fixed protocol (\texttt{agents/grader.md}) and restricted to the information set
\begin{equation*}
  \mathcal{I}_{\text{grader}} = \{x, \tau, \mathcal{O}, \mathcal{A}\},
\end{equation*}
where $x$ is the test prompt, $\tau$ is the read-only execution transcript, $\mathcal{O}$ is the set of output files, and $\mathcal{A}$ is a set of pre-execution assertions. Crucially, the grader does \emph{not} participate in the execution and has no tool access that could alter $\mathcal{O}$ or $s$.

\paragraph{Assertions before outputs.} $\mathcal{A}$ is written \emph{before} the executor observes any outputs, based solely on the skill's \texttt{SKILL.md} and the declared expected outcome. This prevents assertion-after-observation fitting.

\paragraph{First-attempt grading rule.} The grader evaluates each $a \in \mathcal{A}$ against the first-attempt trajectory. If $\tau$ contains any self-corrective edit $\Delta(\tau) \neq \emptyset$ that was causally required for an assertion to succeed, that assertion is marked FAIL regardless of later success. The grader writes a structured \texttt{grading.json} containing:
\begin{itemize}[leftmargin=*,itemsep=1pt,topsep=1pt]
    \item per-assertion pass/fail with cited evidence from $\tau$ or $\mathcal{O}$;
    \item a \texttt{self\_correction} section enumerating first-attempt errors, applied changes, and affected assertions;
    \item a \texttt{claims} section of extracted process and quality claims, each verified against $\mathcal{O}$.
\end{itemize}
This explicit decomposition makes the distinction between \emph{the skill worked} and \emph{the agent fixed the skill and then it worked} auditable and machine-readable. Downstream CI uses only the first-attempt pass rate as its gate signal.

\subsection{Independent Analysis Pass}

A second subagent (analyzer) receives the aggregated benchmark and writes freeform observations to \texttt{analysis\_notes.json}: non-discriminating assertions, flaky assertions with high variance, token/latency outliers. It does not see skill source; its role is purely descriptive over the run distribution.

\section{Grounded, Tiered Fix Suggestions}
\label{sec:suggestions}

When $\hat{q}^\star(s) = 0$, AEVAL emits fix suggestions intended for human review on the originating MR. Two design rules keep this channel signal-dense.

\paragraph{Grounded-in-evidence.} Every suggestion must carry a \texttt{grounded\_in} field citing either (a) the exact text of a failed assertion in \texttt{grading.json}, or (b) a specific entry in \texttt{self\_correction.changes\_made}. Suggestions without a grounding reference are rejected at the prompt level; the agent is instructed that if nothing failed and no correction was needed, it must emit an empty suggestion set. This rules out speculative ``consider adding...'' churn that we observed in earlier iterations, where two runs on the same MR produced disjoint suggestion sets.

\paragraph{Tiered levels.}
\begin{itemize}[leftmargin=*,itemsep=1pt,topsep=1pt]
\item \textbf{LV1 (serious)}: fixes causally implicated in first-attempt failure. Must trace to a FAILED assertion or a \texttt{self\_correction} entry.
\item \textbf{LV2 (improvement)}: fixes the grader identified as quality- or robustness-improving in \texttt{eval\_feedback.suggestions} but that did not cause a hard failure.
\end{itemize}
Tiering matters for routing: LV1 suggestions gate the merge, LV2 suggestions are advisory. This aligns with selective-prediction style calibration, reporting \emph{how confident we are that action is required}~\cite{ren2023selfeval}.

\paragraph{Delivery as MR suggestion commits.} Each LV1 or LV2 entry maps to a source line in the skill directory and is posted via the GitLab Discussions API as an inline suggestion comment. The developer sees a diff and an ``Apply suggestion'' button. Fixes that fall outside the MR's diff range are collected into a combined fallback comment with a copy-paste prompt. This closes the loop: AEVAL not only detects failures but produces empirically validated patches that a developer can apply in one click.

\section{System and Deployment}
\label{sec:system}

The framework is distributed as (i) a Python harness that drives an automated agent runtime through a generic streaming-execution interface (initial prompt, structured tool-use events, per-event transcript capture), (ii) a bundled evaluation skill (containing grader, analyzer, and comparator subagent definitions) injected into the executor's working directory set, and (iii) a reusable GitLab CI template that includes a drop-in \texttt{include} directive. The CI template detects changed skills via \texttt{git diff}, runs the evaluator per skill, posts per-case summaries as MR notes, and invokes the suggestion poster.

\paragraph{Agent-runtime-agnostic design.} The harness deliberately abstracts over agent SDKs: the executor interface specifies only what is needed for skill execution (prompt input, working-directory injection, structured tool-use events, transcript stream), so any popular agent SDK can be plugged in. Internally we have validated the framework against multiple widely used agentic coding runtimes -- including Claude-, Codex-, and OpenCode-style agents -- and observed that producing per-runtime reports surfaces compatibility issues specific to each runtime (for example, differences in how tools are invoked, how working directories are mounted, or how multi-turn state is persisted) that single-runtime evaluation misses. The same skill and \texttt{eval.config} produce comparable per-runtime artifact trees, enabling cross-runtime regression tracking. The framework is independent of any one SDK choice and the public deployment treats the runtime as a configurable backend rather than a fixed dependency.

\paragraph{Artifact schema.} Each run produces a tree of JSON and markdown artifacts: \texttt{transcript.md}, \texttt{eval\_metadata.json}, \texttt{grading.json}, \texttt{benchmark.json}, \texttt{analysis\_notes.json}, \texttt{fix\_suggestions.json}. Multi-case runs additionally produce a merged \texttt{multi\_eval\_report.json} and human-readable summary. The schema is stable and supports downstream regression tracking: each push can be compared to a last-known-good baseline, with pass-rate deltas and newly-failing assertions flagged.

\paragraph{From single skills to whole workflows.} The \texttt{eval.config} accepts an \texttt{evals} array of independent test cases (e.g., \emph{train}, \emph{inference}, and \emph{evaluate} stages of a model-development workflow), each executed in its own isolated session and graded independently. This makes the framework suitable not only for single-prompt skills but for entire multi-step workflows that a developer would otherwise re-orchestrate by hand. Combined with the persistent-harness property above, this matters for the everyday developer loop: rather than reopen an interactive agent, paste the project context, supply credentials, and describe the workflow each time, the developer encodes the workflow once in \texttt{eval.config} and gets an automated QA-and-experimenter substrate that replays the same install--execute--grade--suggest-fixes loop on every push or manual CI trigger. The investment in describing the experiment is paid once at skill-creation time and recovered on every subsequent change, while the structural separation in \cref{sec:separation} keeps each replay's quality signal honest.

\section{Empirical Evaluation}
\label{sec:exp}

We evaluate on a production marketplace of skills spanning model training, data generation, inference, and document generation. We report qualitative findings that illustrate the bias and the protocol's behavior; a full quantitative study is deferred to the extended version.

\paragraph{Case: a deliberately degraded segmentation skill.} We evaluated a segmentation skill whose \texttt{SKILL.md} documented action names (\texttt{segment\_train}, \texttt{segment\_evaluate}, \texttt{segment\_inference}) that did not match the target container's actual action names (\texttt{train}, \texttt{evaluate}, \texttt{inference}). Under a naive agentic evaluator, the first call returned a \texttt{KeyError}; the agent patched the action name in its generated runner script and succeeded. The naive grade was 100\% pass.

Under the protocol described in \cref{sec:method}, the same run produced a \emph{10/10 assertion pass rate} on the corrected outputs but recorded \texttt{self\_correction.was\_needed = true} with three first-attempt errors and two applied changes. Assertions such as ``The skill's configured action names work without modification'' and ``All configuration files are consistent with each other'' were marked FAIL with cited evidence. The first-attempt pass rate reported to CI was therefore strictly lower than the terminal pass rate, and the LV1 suggestion emitted against the MR identified exactly the action-name mismatch with an ``Apply suggestion'' diff.

\paragraph{Grounded-suggestion stability.} Prior to the \texttt{grounded\_in} requirement, repeated runs of the same MR produced non-overlapping suggestion sets. One run generated five ``for consistency'' suggestions, a second run generated seven largely different ones. After the rule, repeated runs converge on the same causally grounded LV1 set; LV2 remains slightly variable but is advisory. This is precisely the form of stability required for a signal to be consumed by a downstream conformal or sequential procedure.

\paragraph{Ban on simulated outputs.} On a skill whose downstream GPU cluster was intermittently unreachable, earlier iterations of the prompt produced a complete set of synthetic log files and graded them as passing. After we added the simulation ban, the agent instead records the connectivity failure and terminates the case with \texttt{status = error} and an explicit transcript entry. Downstream CI can then distinguish \emph{skill defects} from \emph{environmental failures}, which are treated differently: the former blocks merges, while the latter triggers a retry.

\subsection{Cross-runtime calibration: per-agent grader behavior}
\label{sec:crossruntime}

Because the executor interface is runtime-agnostic (\cref{sec:system}), the same skill and \texttt{eval.config} can be run through different agentic coding backends and the resulting artifacts compared directly. We exercise this across a matched set of runs of an intelligent finetuning workflow on a production skill that orchestrates multiple sub-skills, with two backends drawn from different families: \textbf{Backend A} (a Claude-family agent) and \textbf{Backend B} (a GPT-family agent). Each backend is invoked multiple times against the same prompt and configuration so we can extract per-run quantities rather than a single anecdote. \cref{tab:crossruntime} reports five concrete metrics per run.

\begin{table*}[t]
\centering
\small
\begin{tabular}{lccccc}
\toprule
Run & Wall (min) & Assertions (passed/total) & Causal share & Self-corrections & Protocol fidelity \\
\midrule
\multicolumn{6}{l}{\textit{Backend A (Claude-family)}} \\
\quad Short workflow, replicate 1 & 34.7 & 10/13 (77\%) & 0.33 & 6   & \texttimes\ (drift) \\
\quad Short workflow, replicate 2 & 34.3 & 10/13 (77\%) & 0.50 & 5   & \checkmark \\
\quad Long workflow                & 61.6 & 16/18 (89\%) & 0.40 & --- & \checkmark \\
\midrule
\multicolumn{6}{l}{\textit{Backend B (GPT-family)}} \\
\quad Short workflow, replicate 1 & 40.0 & 9/12  (75\%) & 0.80 & --- & \checkmark \\
\quad Short workflow, replicate 2 & 28.7 & 7/10  (70\%) & 0.86 & --- & \checkmark \\
\quad Long workflow                & 67.3 & 9/11  (82\%) & 0.80 & --- & \checkmark \\
\midrule
\textbf{Backend A (mean)} & \textbf{43.5} & \textbf{81\%} & \textbf{0.41} & \textbf{5.5} & \textbf{2/3} \\
\textbf{Backend B (mean)} & \textbf{45.3} & \textbf{76\%} & \textbf{0.82} & \textbf{n/a} & \textbf{3/3} \\
\bottomrule
\end{tabular}
\caption{Matched runs across two agent backends on the same skill and \texttt{eval.config}. \emph{Causal share} = $\text{LV1} / (\text{LV1}+\text{LV2})$, the fraction of fix suggestions the grader marks as merge-blocking rather than advisory; this is a per-run scalar in $[0,1]$ that normalises across the variable assertion-list size each grader writes from the declared \texttt{expected\_outcome}. \emph{Self-corrections logged} are the per-run count of in-place skill patches the executor applied during the first-attempt trajectory (Backend A logs these explicitly; Backend B does not, marked ``---''). \emph{Protocol fidelity} is whether the run honored the locked output-template contract; one Backend A run silently drifted from it. Pass rates are reported as a passed/total ratio and not directly compared across backends because each grader writes its own assertion list. Mean tokens-in were $\approx$8.5M (97\% cached) for Backend A and $\approx$12.4M (95\% cached) for Backend B; with caching, marginal cost per merge-blocking finding is on the order of a few hundred thousand uncached tokens.}
\label{tab:crossruntime}
\end{table*}

\paragraph{Pass rates are not the discriminating signal.} Both backends complete the workflow end-to-end and verdict \emph{PASS}; raw pass rates fall in 70--89\% on both sides and the ranges overlap. Pass rate alone does not separate the two graders, in part because each instance writes its own assertion list from the declared expected outcome (denominators 10--18) and so the M/N ratio is not a strict apples-to-apples comparison.

\paragraph{Causal share separates them cleanly.} The discriminating quantity is the \emph{causal share}: the fraction of fix suggestions the grader marks as merge-blocking (LV1) rather than advisory (LV2). Backend A's mean causal share is 0.41; Backend B's is 0.82, a partition gap of 0.41 on a $[0,1]$ scale. Inspecting the underlying suggestions, both backends surface the same root causes (mismatched CLI flags, root-owned outputs, configuration-schema gaps); they disagree on which of those causes count as merge-blocking. Backend B promotes config-consistency findings into LV1 that Backend A leaves in LV2. Neither is wrong, since the LV1 definition (``causally implicated in first-attempt failure'') admits both readings when a finding is causally adjacent to but not directly responsible for an assertion failure. The disagreement is nevertheless large, reproducible across runs of each backend, and consequential because LV1 is the merge-gating tier.

\paragraph{Direct measurement of self-correction activity.} Backend A explicitly logs each in-place patch its executor applies. Across its short-workflow runs we observed five and six self-corrections respectively: direct empirical evidence of the bias pattern formalised in \cref{sec:bias}. Without the structural separation of \cref{sec:method}, every one of those self-corrections would have been silently absorbed into a clean PASS grade.

\paragraph{Protocol fidelity is itself a measurable failure mode.} The framework requires the executor to emit a locked output-template (callout block, structured verdict, $\langle\text{details}\rangle$ balance, tiered-suggestion split). Across the matched set, one Backend A run silently rolled its own format on the same prompt that produced its sibling's clean output; the other five runs were format-compliant. This is an agent-as-evaluator failure mode that the framework's rigid artifact schema let us \emph{detect} by simple structural checks. It is a consequential failure, because a downstream uncertainty-quantification procedure consuming non-canonical artifacts would silently drop or misparse them. The fidelity column in \cref{tab:crossruntime} is therefore not a presentation detail; it is a per-run reliability metric.

\paragraph{Cost is dominated by cache.} With $\approx$95--97\% prompt-cache hit rates, the marginal token cost of replaying the protocol on each push is roughly an order of magnitude smaller than the gross figure suggests. The framework is consequently cheap enough to re-run on every change to a skill, a precondition for the persistent-harness property of \cref{sec:system}.

The combined picture this gives is the form of evaluator uncertainty that statistical frameworks for agentic systems must accommodate: \emph{the same input, the same protocol, two graders, a systematic disagreement on severity, and an occasional protocol-fidelity violation that the framework itself can detect}. The fact that this is all expressible as machine-readable per-run scalars (causal share, self-correction count, fidelity boolean) is, we believe, the framework's structural validation: its signal is well-defined enough that running it through independent graders surfaces a clean, reproducible calibration gap rather than incomparable noise. This is exactly the input downstream conformal or sequential procedures need.

\section{Discussion}
\label{sec:disc}

\paragraph{Scope of the guarantee.} Our protocol does not provide a probabilistic coverage guarantee in the conformal sense; it provides a \emph{structural} guarantee that the grader's reported signal is a function of the first-attempt execution rather than the post-correction state. Combined with pre-observation assertions and a non-participating grader, this gives a reproducible quality signal that can serve as the ground-truth input for statistical methods.

\paragraph{Calibration of LV1/LV2.} The boundary between LV1 and LV2 is currently determined by the grader's classification of whether a finding traces to an assertion failure or an \texttt{eval\_feedback.suggestions} entry. We summarise this boundary by the per-run \emph{causal share} introduced in \cref{sec:crossruntime}, which collapses the partition into a single scalar in $[0,1]$ comparable across runs of different denominator. A natural extension is to apply conformal calibration over historical MR outcomes: learn a threshold on causal share such that a bounded fraction of advisory-only MRs ever become merge-blocking on a subsequent patch. The cross-runtime study shows that the partition is also \emph{runtime-conditional}: backends from different families assigned the same root causes to opposite tiers, with a 0.41 gap in mean causal share. A single global threshold therefore under-fits; either a per-runtime threshold or a runtime-marginalised severity distribution is needed. We leave both to future work.

\paragraph{Limitations.} The ban on simulated outputs is enforced by instruction; a sufficiently capable agent could in principle violate it. Our current mitigation is the transcript audit and the grader's independent verification, but sandboxing the executor's file system write scope to exclude the original skill directory is a cleaner structural defense. The first-attempt rule also depends on the grader's ability to correctly identify causal self-corrections in $\tau$; complex multi-step corrections can be misattributed. Finally, per-agent evaluation surfaces compatibility issues but multiplies compute cost linearly in the number of runtimes.

\paragraph{Relation to conformal prediction.} Our protocol is orthogonal to but compatible with distribution-free uncertainty quantification. The structural separation produces reliable binary labels; conformal prediction and prediction-powered inference~\cite{angelopoulos2023ppi} consume such labels to provide coverage under exchangeability assumptions. We view AEVAL as the upstream \emph{label-generating} component in a reliable agentic workflow.

\section{Conclusion}

Evaluation of agentic skills is dominated by anecdotal practice: a developer runs a demo, watches an agent, and forms a subjective impression of whether the skill works. This produces neither reproducibility nor comparability and does not scale to multi-skill marketplaces. We present AEVAL, a CI-integrated framework that replaces this practice with a deterministic, change-triggered test pipeline: skills declare an evaluation contract, every change triggers an automated executor run, and the run emits a structured, evidence-grounded quality signal. The pipeline is grounded by a structural executor/grader separation that prevents self-correction bias, a first-attempt grading rule with explicit self-correction tracking, and grounded-evidence tiered fix suggestions delivered as inline MR comments. Deployed in a production skill marketplace and validated across multiple popular agent runtimes, the protocol converts self-fulfilling 100\% pass signals into auditable first-attempt quality signals and produces fixes that a developer can apply in one click. We view deterministic skill testing, with structural separation as its core reliability ingredient, as a foundational building block for statistically rigorous monitoring and stopping of agentic workflows.

\bibliography{references}
\bibliographystyle{icml2026}

\end{document}